# INDIRECT OBJECT REPRESENTATION AND ACCESS BY MEANS OF CONCEPTS

Alexandr Savinov

*Institute of Mathematics and Informatics, Academy of Sciecnces of Modova, str. Academiei 5, 2028 Kishinev, Moldova*
*savinov@conceptoriented.com, http://conceptoriented.com/savinov*

Keywords: Indirect object representation and access, Concept, Concept-oriented programming, Separation of concerns.

Abstract: The paper describes a mechanism for indirect object representation and access (ORA) in programming languages. The mechanism is based on using a new programming construct which is referred to as concept. Concept consists of one object class and one reference class both having their fields and methods. The object class is the conventional class as defined in OOP with instances passed by reference. Instances of the reference class are passed by value and are intended to represent objects. The reference classes are used to describe how objects have to be represented and accessed by providing custom format for their identifiers and custom access procedures. Such an approach to programming where concepts are used instead of classes is referred to as concept-oriented programming. It generalizes OOP and its main advantage is that it allows the programmer to describe not only the functionality of target objects but also intermediate functions which are executed behind the scenes as an object is being accessed.

## 1 INTRODUCTION

In OOP all references have the same format and all objects are being accessed using one and the same procedure. However, these peculiarities are hidden and the programmer has the illusion of *instant* or *direct* access as if the action started immediately after the method was called. (It is analogous to the action-at-a-distance principle in classical physics.)

In this paper it is assumed that object access is always *indirect*. This means that object interactions are mediated by other objects playing a role of spaces or environemnts. The format of references can be specified by the programmer and there is always some intermediate procedure implementing the underlying logic of access. In this sense the main goal of this approach consists in providing language means for *indirect* object representation and access (ORA) in programming languages. For example, let us consider a simple method invocation: `myObject.myMethod()`. In OOP reference `myObject` has some standard format that cannot be changed. Using the proposed approach it is possible to define any appropriate format of reference `myObject` in the program where the object is used. For example, this reference might consist of two fields: a unique integer and the time of creation. Or it might contain a computer name, port and a unique object identifier. The target method `myMethod` will be wrapped into an intermediate reference resolution procedure which is also is written by the programmer. However, such a customization of access retains the illusion of instant access, i.e., we still apply the target method to a reference and do not care about intermediate actions.

Currently there exist many different ORA mechanisms provided by operating systems, middleware or standard libraries. For example, there exist numerous standard implementations for such ORA mechanisms as local heap, global heap, remote objects, persistent objects, managed objects, transactional objects etc. Each of them uses its own format of references such as 16- or 32-bit integers, table name and primary key, computer name and object identifier. Accordingly, these references are served by their own access procedures which however do not belong to the application program. Depending on the needs the programmer can create program objects in one of these standard containers. For example, frequently created and deleted small objects could be placed in the local heap. Objects intended to be passed over the network are created as remote objects. And objects that need to have a longer life-cycle can be created in the persistent container.

So what is wrong with this traditional approach? The main problem is that frequently we want to develop our own custom ORA mechanism designed for the purposes of this concrete program. In this case using some standard library or middleware does not help because it is not integral part of the program and cannot be (easily) changed and adapted to its needs. A solution in this case consists in integrating the mechanism of ORA into the program itself at the level of the programming language. Such a program is dealing not only with objects themselves but also how they are represented and accessed. Hence it can be viewed as consisting of two types of functionality: (i) normal *business methods* (BMs) called explicitly by the programmer, and (ii) *object representation and access* (ORA) methods called implicitly behind the scenes. In other words, ORA methods constitute hidden part of the program overall functionality because their calls cannot be found in the source code. In the existing approaches the language is designed to describe only BMs while the hidden part is separated from the main program and belongs to the operating system level, middleware or a library. In the proposed approach this hidden ORA functionality is integral part of the program in the sense that it is described along with the objects using appropriate languages constructs. For example, the programmer might develop a custom local heap, global heap or any other type of container implementing one or another ORA strategy which specific to this system.

A program consisting of the two types of functions can be viewed and designed as a structured space described in Section 2. In order to define how program objects have to be represented and accessed we propose a new language construct, called concept, which is described in Section 3: In particular, this new construct allows the programmer to implement the following two mechanisms: reference substitution and resolution is described in Section 4 and reference concatenation is described in Section 5. Section 6 describes some related work and Section 6 provides concluding remarks.

## 2 OBJECTS IN SPACE

A program in the described approach can be viewed as a set of nested spaces where objects live (Fig. 1). Each space has one parent space and a number of child spaces or objects. This space structure is precisely what is referred to as *physical structure* in the concept-oriented data model (Savinov, 2006). According to this analogy objects can interact only by intersecting the intermediate space borders. Any access request such as a method call or message cannot *directly* (instantaneously) reach its target object. Instead, it follows some access path starting from the source context and leading to the target context. Each object is a unique position within the structured space which is stored and passed via references. In order to access an object it is necessary to get its reference and then to interpret it as an address in the space. This is precisely why such an access method is referred to as *indirect*. Each intermediate border along the access path is actually a normal object with special functions for intermediate processing. These intermediate functions are triggered automatically as some access request intersects this border. The target method is only the last step in this indirect access procedure and frequently it accounts for a relatively small portion of the overall system complexity. In large systems most of their functionality is concentrated on intermediate space borders. Notice that these functions are hidden because they are never called explicitly in the program. For example, if a Java method is called then the target reference is resolved into a memory handle, which is then resolved into a physical address, which in turn is somehow processed by CPU and finally is passed to the hardware memory manager. Notice that all these actions are executed behind the scenes and are not part of the program. In this sense the goal of the proposed approach consists in proving support for describing such a structured space at the level of the programming language.

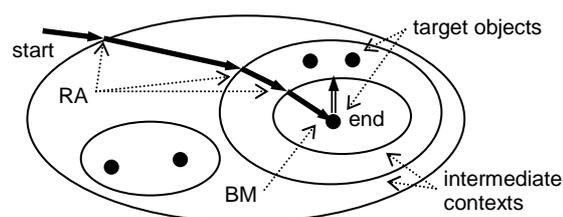

Figure 1: Program structure.

Each space is responsible for ORA to its children. In particular, it determines the format of its own local identifiers used to distinguish internal elements. Whenever an internal object needs to be accessed, its space has to resolve the local identifier. Such a structure is analogous to a hierarchical coordinate system. For example, postal address is specified as a country name, city, street and house number. Countries are specified in the outer most space; then each country has a number of cities and so on. In the described approach we assume that objects in the program have the same hierarchical structure.

Any hierarchical coordinate system has two mechanisms: (i) reference substitution (Section 4), and (ii) reference concatenation (Section 5). Reference substitution allows us to introduce new object identifiers which substitute existing identifiers. For example, a computer name (DNS) substitutes for some IP address, which in turn substitutes for some physical (MAC) address. Java reference is a substitute for some memory handle, which in turn replaces some physical address in memory. This mechanism allows us to bring a new level of indirection and unbound objects identifiers from the reality. However, in order to access such an indirectly represented object in future we need to resolve its reference. This process of resolution proceeds infinitely deeply in the physical structure. For example, an object physical address is somehow processed by the memory manager, which translates it into commands for the memory chip, which then transforms these commands to analogue impulses and so on. However, we always can (and should) choose some level of representation which is considered final and ultimate. In other words, we assume that there is some special identification system, called *root*, all our references are resolved into. Normally the root space is provided by the compiler and depends on the platform. It is precisely what is used in OOP to directly representing *all* objects.

The dual mechanism of reference concatenation allows us to create hierarchical references from several segments. Each next segment in such a complex reference identifies the next space or object relative to the previous segment. For example, one postal address concatenates several segments such as country, city, street and house. Each segment describes a context for the next segment. Domain name system has also a hierarchical structure where segments are concatenated. For example, www.icsoft.org consists of three segments with high segment org followed by the second segment icsoft and ending with low segment www.

Thus any object in the structured space and any object in the proposed approach are identified by a complex reference consisting of several segments. Each segment is a position of this object relative to the parent object (concatenation). On the other hand, each segment substitutes for and needs to be resolved into some root reference which can be then used for direct access. For example, a persistent object could be identified by three segments: a database name, a table name and a primary key. In order to access such an indirectly represented object we need to intersect three borders each of which triggers some functionality. In particular, each of these segments has to be resolved into its root reference which has been substituted.

## 3 CONCEPT DEFINITION

A hierarchical coordinate system described in the previous section can be modelled using a new programming construct, called *concept*. Concept is a pair consisting of two classes: an *object class* and a *reference class*. The object class is a normal class as used in OOP (so concept without a reference class is equivalent to class as defined in OOP). Its instances, called *objects*, are passed by means of references. Instances of the reference class are passed by value and are intended to represent objects. Thus an object reference is its address which can be stored and passed by value to other objects as its representative.

It is important that concept is by definition a dual construct and its two constituents are intended to describe simultaneously two types of functionality. Notice that they can be separated only in special cases where a concept with the empty reference class is a normal class while a concept with the empty object class describes some reference format or an object passed by value. Concepts are used instead of classes to declare a type of variables, parameters, fields and other elements of the program. An approach to programming based on using concepts is referred to as *concept-oriented programming* (Savinov, 2005).

Listing 1: Concept definition.
```
01  concept MyConcept in ParentConcept
02    class {
03      double objField; // Passed by ref
04      void continue() { ... }
05      int myMethod() { ... }
06    }
07    reference {
08      int refField; // Passed by value
09      void continue() { ... }
10      int myMethod() {
11        ...
12        double tmp = context.objField;
13        ...
14      }
15    }
```

Listing 1 is an example of concept definition. Its object class (lines 2-6) has one double field (line 3),

one special method `continue` (line 4) and one normal method `myMethod` (line 5). The reference class (lines 7-15) has one integer field (line 8) intended to identify other objects, one special method `continue` intended to resolve this reference (line 9), and one normal method `myMethod` (lines 10-14). Notice that one concept may have two definitions for one method, which are called *dual methods*. Once we have defined some concept it can be used as a type. For example, we might define a field or a variable as follows: `MyConcept myVar;` After that this variable can be used as usual for method invocation: `myVar.myMethod();` However, in contrast to OOP, this variable may store rather complex data which is the address of this object. The object itself may reside anywhere in the world and be accessed via rather complex procedure executed behind the scenes after this method is called.

In terms of the hierarchical coordinate system described in the previous section an object of a concept is one space and a reference of a concept is one coordinate within this space. Normally one concept produces many objects and each of these objects has many references representing internal objects. An object where this reference has been created is referred to as *context*. Thus an important distinguishing feature of CoP is that any element of the program lives in the context of some other object. Access to the current context in the program code is provided by the keyword `'context'`. If this keyword occures in some reference method then it provides access to the object of this same concept where this reference has been created. For example, reference method `myMethod` in Listing 1 (lines 10-14) uses this keyword in order to get a double value stored in the object of this concept (line 12).

Each reference is intended to identify one internal object within its context. Nested space structure is modelled via concept *inclusion relation*, which means that any concept has to specify a parent concept in its declaration. For example, concept `MyConcept` in Listing 1 is included in `ParentConcept` using keword `'in '` (line 1) As a consequence objects of `MyConcept` will be represented by means of references of `ParentConcept`. Objects and references exist at run-time within a hierarchy having the structure defined by the concept inclusion relation. A property of this hierarchy is that objects and references interleave so that each reference represents some object which has a number of its own references which again represent some objects and so on. For example, in Fig. 2 objects and reference are denoted by white and grey rectangles, respectively. Arrows denote 'representd by' relationship between objects and references. One root object X has two references which represent two objects (reference A represents object Y). Each of these two objects also has two references representing target objects without references (reference B in context Y represents object Z). Any object is a context for all elements which are positioned below.

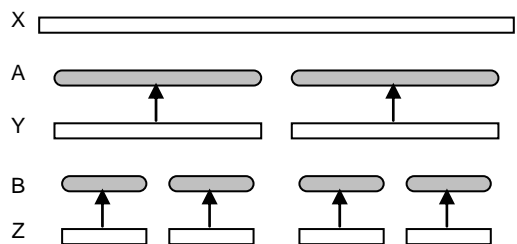
Figure 2: Object and reference run-time structure.

An object is represented by its *complex reference*, which is a sequence of segments where each segment is one parent concept reference. The first segment is of root concept (the root of the concept hierarchy). The second segment is of the next child concept leading to the target object and so on. Each intermediate segment represents one context. For example, object Z in Fig. 2 is represented by a complex reference consisting of two segments A and B: <A,B> The first high segment A represents context Y while the second low segment B represents context Z which is the target object. Notice that there is always some initial well known context we start from, which needs not to be represented like X in this example.

A concept has two special *continuation methods* defined in its object class and reference class (lines 4 and 9 in Listing 1). They play a very important role in the ORA mechanism by providing a door in the space border described by this concept. The reference continuation method allows us to pass through this reference to the parent reference. The object continuation method allows us to pass through this object to the parent object. In this paper we consider only the reference continuation method the main role of which consists in resolving this reference. Given a complex reference the ORA mechanism has to resolve its segments into the root references which are used for *direct* access to the represented objects.

## 4 REFERENCE SUBSTITUTION

The mechanism of reference substitution and resolution is intended for creating a new level of indirection where a new format of reference is used to represent objects instead of existing references. Whenever an object is going be accessed this new reference has to be resolved into the substituted reference. In CoP each new child concept in the inclusion hierarchy can define its own reference which then substitutes for a parent reference. Thus each new child concept can be used as a new level of indirection by introducing a new format of references. For example, in Fig. 3 reference C indirectly represents some target object by substituting for reference B, which in turn substitutes for reference A of the root type. Accordingly, in order to access such an object it is necessary to resolve reference C into A and only after that to execute the target method specified in the source context. Thus the target method is wrapped into the resolution procedure with the sequence shown by arrows in Fig. 3.

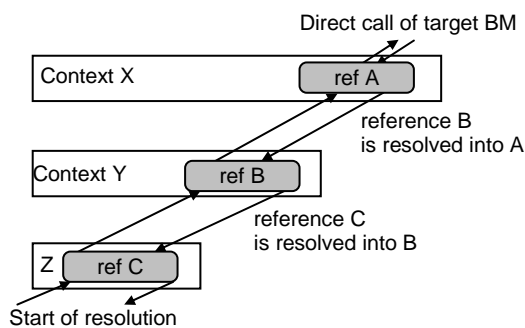

Figure 3: Reference substitution and resolution.

An example in Listing 2 illustrates the logic of reference substitution and resolution. The general goal is that we want to represent our objects indirectly by using our own custom references but at the same time retaining the illusion of instant access, which means that we apply methods as usual and all the necessary intermediate operations are carried out seamlessly behind the scenes. More specifically, the goal consists in developing a very simple persistent storage mechanism where objects of arbitrary target classes are identified by their integer primary keys. The logic of ORA is described in concept Persistent. It has one static field (line 3) in its object class, which is a reference to a real persistent storage like database.

In order to uniquely identify objects in the context of the persistent storage we define a reference class with one integer field (line 6) which contains a value of the primary key for the object of internal class included in this concept. The method of continuation of the reference class (lines 7-13) is intended for resolving this reference into the substituted root reference. This method uses the integer field as a key and loads the target object from the storage (line 9). Here the direct root reference to the target object is restored and the continuation method applies the next continuation method (line 10). This effectively means that the compiler will pass control to the restored object and the requested business method will be executed. When access is finished the state of the target object is stored back in the database (line 11).

Notice that the reference resolution method (lines 7-13) is unaware of the target object type and the business method being invoked. It simply gets control from somewhere, resolves the reference (line 9), then passes control further to the just resolved object (line 10), and finally stores the objects before it returns (line 11). On the other hand target objects and their business methods do not involve the logic of indirect ORA. Thus we effectively separated two concerns: BMs and ORA:

Listing 2: An example of reference substitution.
```
01 concept Persistent
02   class {
03     static Storage st.create();
04   }
05   reference {
06     long id; // Primary key
07     void continue() {
08       print("> Start of resolution\n");
09       Root r = context.st.load(id);
10       r.continue();
11       context.st.store(id, r);
12       print("< End of resolution\n");
13     }
14   }
```

Let us now consider how this concept is used (Listing 3). Assume that a new class Account has been developed (lines 1-11), which implements some business logic specific to exclusively account management such as debiting, crediting, getting/setting balance etc. Accounts have to be persistent objects. However, we do not want to include any logic of ORA into this class. The reason is that in future the logic of ORA or the logic of account management might well be changed and therefore these two concerns should be placed in different modules. In order to solve this problem class Account is included in concept Persistent using keyword 'in' its declaration (line 1).

Because of this inclusion the compiler will generate references to account objects as having the format of its parent concept, i.e., each account will be automatically represented by an integer. When the account is going to be accessed the compiler will automatically wrap the target business method into the resolution procedure provided by the corresponding reference class. Thus we reached our design goal: accounts are represented indirectly by integers but can still be used as if they were normal objects. For example, method `credit` in class `Source` gets an account reference as a parameter (line 14). Here again, this parameter has the format specified by concept `Persistent` which is a parent of class `Account`. In order to credit the account we get its balance (line 15), then add the specified amount (line 16) and finally save the new balance back in the account object (line 17). However, each target account method invocation will be executed indirectly using the intermediate reference resolution procedure. Method `credit` (line 14-18) will generate the following output:

```
$ > Start of resolution
$   * getBalance is called
$ < End of resolution
$ > Start of resolution
$   * setBalance is called
$ < End of resolution
```

Here we see that both BM invocations are wrapped into their reference resolution procedures which print a pair of lines around each target method.

Listing 3: An example of indirect ORA.
```
01 class Account in Persistent {
02   double b = 0;
03   double getBalance() {
04     print("  * getBalance is called\n");
05     return b;
06   }
07   void setBalance(double t) {
08     print("  * setBalance is called\n");
09     b = t;
10   }
11 }
12
13 class Source {
14   void credit(Account a, double m) {
15     double t = a.getBalance();
16     t += m;
17     a.setBalance(t);
18   }
19 }
```

An advantage of this procedure is that two types of functionality are separated in a principled manner: the logic of ORA is described in concept `Persistent` while the business logic is described in class `Account`. The use of the indirectly represented objects is completely transparent because we are actually unaware of ORA mechanism used to access the target account object in class `Source`. Another advantage is that we can develop multiple levels of indirection where each new level provides references substituting for previous (parent) references. For example, if accounts have to be represented by their unique names then a new concept can be defined which is included into concept `Persistent`. Its reference class will have one string field which substitutes for some integer which in turn substitutes for a root reference. Notice that the source code where account objects are used does not depend on all these changes. We simply use account objects as if they were represented and accessed directly via root references while the intermediate code is injected automatically by the compiler according to the concept inclusion hierarchy.

## 5 REFERENCE CONCATENATION

Reference substitution described in the previous section is intended for modelling a nested structure of indirection. The dual mechanism is referred to as *reference concatenation* and its purpose consists in modelling a hierarchical address system. The idea here is that an address (coordinate, reference, identifier etc.) consists of several segments each of them representing the next subspace and the last segment representing the target object. This sequence of segments is referred to as *complex reference*.

In the example shown in Listing 2 it was assumed that the object class does not produce instances because it has only static fields (line 3). In this case there exists only one well known persistent storage at run-time. However, normally concepts will contain non-static fields and hence their object classes will be used to produce many instances. If concept `Persistent` has a non-static field in its object class then it is not enough to store an integer in order to uniquely identify a target object. In addition to this integer we have to remember the object where it has been created as the first segment. In other words, a complete reference has the database identifier as the first segment and the primary key within this database as the second segment. Such a hierarchal reference structure is also described by the concept inclusion relation. As usual an object is represented by its parent concept reference. However, since this reference exists in some context we need to represent also this context by using its own parent reference and so on till the root.

A program in Listing 4 is a modified example of Listing 2. The main difference is that concept `Persistent` has a non-static field referencing a persistent storage (line 17). Objects of concept `Persistent` will be represented by references provided by its parent concept `NamedObjects`. Pbjects of classes included in `Persistent` (such as accounts) will be represented by complex references consisting of two segments: database string identifier (line 6), and an integer primary key (line 21). Since any reference has two segments the access procedure consists of two steps. On the first step we need to resolve the first segment and enter the first context. On the second step it is necessary to resolve the second segment and enter the second context. The first segment is resolved by the reference continuation method of concept `NamedObjects` (lines 7-12). This method simply looks up the name in the map (line 9) and then proceeds (line 10). After that the second segment can be resolved by calling the reference continuation method of concept `Persistent` (lines 22-34). This method opens the database before access (line 25) and closes it after access (line 32) if no other processes are using it. The rest of the procedure is identical to that in Listing 2.

Listing 4: An example of hierarchical access.

```
01 concept NamedObjects
02   class {
03     static Map map.create();
04   }
05   reference {
06     String id;
07     void continue() {
08       print("> Enter NamedObjects\n");
09       Root r = context.map.get(id);
10       r.continue();
11       print("< Exit NamedObjects \n");
12     }
13   }
14
15 concept Persistent in NamedObjects
16   class {
17     Storage st.create();
18     int accessCount = 0;
19   }
20   reference {
21     long id; // Primary key
22     void continue() {
23       print("  > Enter Persistent\n");
24       if(context.accessCount++ == 0);
25         context.st.open();
26       accessCount++;
27       Root r = context.st.load(id);
28       r.continue();
29       context.st.store(id, r);
30       accessCount--;
31       if(context.accessCount == 0);
32         context.st.close();
33       print("  < Exit Persistent\n");
34     }
35   }
```

Let us now again consider how this modified concept will be used using source code from Listing 3. Notice that neither class `Account` nor class `Source` where it is used have to be adapted to the modified ORA mechanism of concept `Persistent`. It is a consequence of our design where ORA logic is separated from business logic. In other words, we still can use class `Account` as if it was represented directly. What has really changed is the intermediate ORA procedure executed behind the scenes. In particular, the account object passed as a parameter to method `credit` (line 14) will be now represented by longer references consisting of two segments. In order to invoke an account method it is necessary to resolve the database name, then (in its context) resolve the object primary key and finally call the method using the object direct (root) reference. Thus method `credit` (line 14-18) will produce the following output:

```
$ > Enter NamedObjects
$   > Enter Persistent
$ * getBalance is called
$   < Exit Persistent
$ < Exit NamedObjects
$ > Enter NamedObjects
$   > Enter Persistent
$ * setBalance is called
$   < Exit Persistent
$ < Exit NamedObjects
```

## 6 RELATED WORK

OOP provides means for describing object functionality using classes and inheritance relation. CoP adds means for describing how objects are represented and accessed using concept as a generalization of class. One method for implementing indirect ORA in OOP consists in using proxies. Proxy is a new class substituting for some target class and explicitly used instead of it. A disadvantage of this approach is that we need to explicitly use proxy class when we want to work with the target class. In contrast, in CoP the programmer always works with the target class and its methods as if its objects were directly accessible while all the intermediate elements are injected automatically. A code with proxies is difficult to maintain because any change in the target class has to be *manually* reflected in the proxy. Another disadvantage is that it is even more difficulty to implement nested proxies. And the third problem is that proxies do not allow us to implement custom reference format.

One method for implementing custom references in C++ consists in using so called smart pointers (Stroustrup, 1991). However, it is a rather specific technique based on using templates, which requires a relatively high degree of manual support.

Aspect-oriented programming (AOP) (Kiczales et al., 1997) allows the programmer to inject any code in special points of the program. AOP introduces an additional construct, called aspect, which exists along with classes. In contrast, CoP introduces concepts which *generalize* classes. However, like proxies, AOP does not provide any means for modelling custom format of references.

Aspects in AOP can be viewed as an analogue of parent concepts in CoP. Both these modules are intended to change the behaviour of target classes. In terms of AOP, both aspects and parent concepts contain some code that is injected in many different target classes. The difference is that in AOP target injection points are declared in the aspect itself and they are unaware of this influence. In contrast, in CoP the target classes themselves specify in their definition what kind of intermediate code has to be injected. The parent concept which contains this intermediate code is unaware of the points where its code is injected.

Another related approach is based on using *mixins* (Bracha et al., 1990). In particular, it is similar to CoP (and AOP) in its ability to wrap some target code into another method (using around keyword). It is a convenient addition to OOP which however fails to solve the problem of modelling references.

CoP is also similar to *context-oriented* direction in programming and layered design (Constanza et al., 2005) because of its ability to put objects in context or environment which changes their behaviour. In ContextL context-orientation is supported by keyword `'in-layer'` while in CoP a class is included in some parent concept via `'in'`.

The same goal of describing custom environment for the program is pursued in the field of reflective middleware and metaobject protocols (Kiczales et al., 1993). In CoP this task is performed by parent concepts which allow the programming to create an environment influencing the behaviour of internal objects.

Concepts may have two definitions for one method (in the object class and the reference class). This property can be used to model super and inner methods of classes (Goldberg et al., 2004).

## 7 CONCLUSIONS

In this paper we described an approach to indirect representation and access in programming languages. Its main distinguishing feature consists in possibility to model references and hidden functionality which is executed during object access. Moreover, it is the only known approach which is directly aimed at modelling the structure of references and their functions (as opposed to the structure of objects and their functions). This goal is achieved by introducing a new programming construct, called concept, which includes a reference class in addition to the conventional object class. An important property of this construct is that it can be used for describing two sides of any program: explicitly invoked business methods and implicitly executed intermediate methods. In particular, we showed how concepts can be used to implement the mechanism of reference substitution and the mechanism of reference concatenation. Such an approach to programming based on using concepts and aimed at describing two types of functionality is referred to as the concept-oriented programming. This approach can be useful for many types of programs but it is especially effective for complex systems where intermediate functionality accounts for a great deal of the system complexity.